\documentclass[letterpage,titlepage,11pt]{article}
\usepackage{mathtools,amssymb,color,epsf,tikz,microtype} 
\usepackage{latexsym,color}
\usepackage{cite}

\usetikzlibrary{decorations.pathreplacing,angles,quotes}
\usetikzlibrary{arrows,decorations.markings}

\usepackage[
      colorlinks=true,
      linkcolor=blue,
      urlcolor=blue,
      filecolor=black,
      citecolor=green,
      pdfstartview=FitV,
      linktocpage,
      pdftitle={},
        pdfauthor={x, x, x},
        pdfsubject={},
        pdfkeywords={},
        pdfpagemode=None,
        bookmarksopen=true
      ]{hyperref}
      
\makeatletter
\@addtoreset{equation}{section}
\makeatother

\allowdisplaybreaks[1]

\begin{document}

\begin{titlepage}
\begin{flushright}
\hfill{ ICCUB 17-011}
\end{flushright} 
\vskip -1cm

\leftline{}
\vskip 2cm
\centerline{\LARGE \bf Non-Relativistic BMS algebra}

\vskip 1.2cm
\centerline{\large\bf Carles Batlle$^{a}$, Diego Delmastro$^{b}$ and Joaquim Gomis$^{b}$}
\vskip 0.5cm
\centerline{\sl $^{a}$Departament de Matem\`atiques and IOC,}
\centerline{\sl Universitat Polit\`ecnica de Catalunya}
\centerline{\sl EPSEVG, Av. V. Balaguer 1, E-08808 Vilanova i la Geltr\'u, Spain}
\centerline{{email: {\tt carles.batlle@upc.edu}}}
\centerline{\sl $^{b}$Departament de F\'{\i}sica Qu\`antica i Astrof\'{\i}sica}
\centerline{\sl and Institut de Ci\`encies del Cosmos,}
\centerline{\sl Universitat de Barcelona, Mart\'i i Franqu\`es 1, E-08028 Barcelona, Spain}
\centerline{{email: {\tt ddelmade7@alumnes.ub.edu}, {\tt joaquim.gomis@icc.ub.edu}}}

\smallskip
\vskip 0.5cm

\vskip 1.2cm
\centerline{\large\bf Abstract} \vskip 0.4cm 
\noindent
{We construct two possible candidates for the non-relativistic $\mathfrak{bms}_4$ algebra in four space-time dimensions by contracting the original relativistic $\mathfrak{bms}_4$ algebra. The $\mathfrak{bms}_4$ algebra is infinite-dimensional and it contains the generators of the Poincar\'e algebra,  together with the so-called \emph{super-translations}. Similarly, the proposed $\mathfrak{nrbms}_4$ algebras can be regarded as two infinite-dimensional extensions of the Bargmann algebra.
We also study a canonical realisation of one of these algebras in terms of the Fourier modes of a free Schr\"odinger field, mimicking the canonical realisation of the relativistic $\mathfrak{bms}_4$ algebra using a free Klein-Gordon field.
}

\end{titlepage}
 
\pagestyle{plain}
\setcounter{page}{1}

\vspace{2mm} \hrule \vspace{1mm} \noindent

\tableofcontents

\vspace{0.75cm}{\vspace{1mm} \hrule \vspace{1mm} \noindent}

\section{Introduction}
Recently there has been a renewed interest in the Bondi-Metzner-Sachs (BMS) group \cite{BMS-1}.  This group is the semi-direct product of the Lorentz group with the infinite dimensional group of super-translations \cite{BMS-1}, which is abelian.
The BMS symmetry plays an important role in the understanding of gravitational scattering, 
soft theorems,  memory effects and black holes; see, for example, the recent lectures by Strominger
\cite{Strominger:2017} on the subject, where one can also find an extended set of references to the original literature. 
Moreover,  the BMS group could play a crucial role  in understanding holography in 
asymptotically 
flat space times \cite{Banks:2003vp}  \cite{deBoer:2003vf}  \cite{Arcioni:2003xx} \cite{Barnich:2010eb}. The BMS symmetry is an infinite conformal extension of the Carroll symmetry \cite{Duval:2014}, which was introduced in \cite{Levy-Leblond} as a limit of the Poincar\'e symmetry when the velocity of light is scaled down to zero.

{
The application of holography to condensed matter systems 
 -- known as non-relativistic holography -- 
is a subject of current interest. Non-relativistic holography has been used to study phenomena such as cold atoms, quantum critical points and high temperature superconductivity \cite{Sachdev} \cite{liu}
\cite{Hartnoll:2016apf}.  Moreover,   the role of non-relativistic gravities to understand strongly coupled systems in condensed matter has been appreciated; 
see, for example, the use 
of Newton-Cartan gravity in the study of the quantum Hall effect 
\cite{Son:2013rqa} \cite{Geracie:2016bkg}.

The study of flat-holography for non-relativistic systems 
 is a useful tool because flat space is a good approximation to the real world. Non-relativistic symmetries, in particular the non-relativistic
super-translations, are expected to be useful in the study of the infrared behavior of non-relativistic quantum field theories, in the same way that the relativistic super-translations allow us to understand soft theorems \cite{Weinberg:1965nx} as Ward identities.
}

In order to analyse this point it is natural to consider the non-relativistic limit of holography in asymptotically Minkowski space-times. With this motivation in mind, in this work we will construct two non-relativistic versions of the $\mathfrak{bms}_4$ ($\mathfrak{nrbms}_4$) algebra and an explicit realization of one of them.

In order to construct the non-relativistic analogue of $\mathfrak{bms}_4$ there are several approaches one could take. The simplest one is to consider a suitable \.In\"on\"u-Wigner contraction of the corresponding relativistic $\mathfrak{bms}_4$ algebra\footnote{The contraction procedure can be extended to an arbitrary number of space-time dimensions\cite{future}.}, in the same sense that the Galilei algebra can be obtained by contracting the Poincar\'e one. An alternative to obtain a possible non-relativistic $\mathfrak{bms}_4$ algebra is to mimic the canonical construction of Longhi and Materassi \cite{Longhi:1997zt}, but using a scalar field with Galilean space-time symmetries instead of a Klein-Gordon field. Finally, a third alternative is to follow the same steps that led to the original $\mathfrak{bms}_4$ algebra
\cite{BMS-1}~\cite{Sachs:1962zza}, but in a non-relativistic setting, i.e., one could try to study the set of isometries of asymptotically flat Newtonian 
space-times~\cite{Cartan:1923zea}, characterized by a contravariant degenerate spatial  metric $h^{\mu\nu}$ and 
covariant vector $\tau_\mu$; see, for example,
\cite{Duval:1984cj} and references therein. Here we will only consider the first two possibilities.

The organisation of the paper is as follows. In Section~\ref{section2} we   will construct the two contractions of the $\mathfrak{bms}_4$ that contain the Bargmann algebra~\cite{Bargmann:1954gh} as a sub-algebra.
In Section~\ref{section3} we  obtain a realisation of one of the $\mathfrak{nrbms}_4$ algebras in terms  of a free Schr\"odinger field. 
Section~\ref{section4} is devoted to conclusions and outlook.

\section{The algebra $\mathfrak{nrbms}_4$ as a contraction of $\mathfrak{bms}_4$}
\label{section2}
In this section we will construct two possible $\mathfrak{nrbms}_4$ algebras as the two possible contractions of the standard
$\mathfrak{bms}_4$ algebra. The $\mathfrak{bms_4}$ algebra is the semi-direct sum of the Lorentz algebra with the generators of the super-translations, which form an infinite-dimensional abelian sub-algebra~\cite{BMS-1}. In a similar fashion, the $\mathfrak{nrbms}_4$ algebra will be given by the semi-direct sum of the Bargmann algebra with the generators of spatial super-translations. The extension of the algebra to include super-rotations~\cite{Barnich:2011ct} will not be considered in this paper.

The canonical realisation of the $\mathfrak{bms}_4$ algebra in terms of the Fourier modes of free Klein-Gordon field \cite{Longhi:1997zt} leads to the  algebra
\begin{equation}
\begin{split}
{}[J_i,J_j]&=i\epsilon_{ijk}J_k,\\
[J_i,K_j]&=i\epsilon_{ijk}K_k,\\
[K_i,K_j]&=-i\epsilon_{ijk}J_k,
\end{split}\hspace{60pt}
\begin{split}
[P_\ell^m,P_{\ell'}^{m'}]&=0,\\
[J_i,P_\ell^m]&=i(\mathcal J_i)_{\ell\, m'}^{\ell'm}\ P_{\ell'}^{m'},\\
[K_i,P_\ell^m]&=i(\mathcal K_i)_{\ell\, m'}^{\ell'm}\ P_{\ell'}^{m'},
\end{split}
\end{equation}
where $J_i,K_i$  are the  generators of Lorentz transformations, and $P_{\ell}^{m}$ are the generators of super-translations. The indices $\ell,m$ are integers such that $|m|\le\ell$, and $(\mathcal J_i)_{\ell\, m'}^{\ell'm},(\mathcal K_i)_{\ell\, m'}^{\ell'm}$ are the structure constants of the $\mathfrak{bms_4}$ algebra. These matrices 
furnish a  non-unitary, reducible but indecomposable infinite-dimensional representation of the Lorentz group (with the sign of the structure constants negated, corresponding to passive vs.~active transformations).

More explicitly, these matrices are given by
\begin{equation}\label{bmsalgebra}
\begin{aligned}
{}[J_1,P_\ell^m]&=\frac{i}{2} \sqrt{(\ell+m) (\ell-m+1)}\ P_\ell^{m-1}-\frac{i}{2}\sqrt{(\ell-m) (\ell+m+1)}\ P_\ell^{m+1},\\
[J_2,P_\ell^m]&=\frac{1}{2} \sqrt{(\ell-m) (\ell+m+1)}\ P^{m+1}_\ell+\frac{1}{2} \sqrt{(\ell+m) (\ell-m+1)}\ P^{m-1}_\ell,\\
[J_3,P_\ell^m]&=m\ P^m_\ell,\\
[K_1,P_\ell^m]&=\frac{1}{2} \sqrt{\frac{2 \ell+1}{2 \ell-1}(\ell-m) (\ell-m-1)}\ P_{\ell-1}^{m+1}\\
&+\frac{1}{2} \sqrt{\frac{2 \ell+1}{2 \ell-1}(\ell+m) (\ell+m-1)}\ P_{\ell-1}^{m-1}\\
&+\frac{1}{2} \frac{(\ell-1) (\ell+3)}{2 \ell+3} \sqrt{\frac{ (\ell-m+1)(\ell-m+2)}{(2 \ell+1) (2 \ell+3)}}\ P_{\ell+1}^{m+1}\\
&+\frac{1}{2} \frac{(\ell-1) (\ell+3)}{2 \ell+3} \sqrt{\frac{ (\ell+m+1)(\ell+m+2)}{(2 \ell+1) (2 \ell+3)}}\ P_{\ell+1}^{m-1},\\
[K_2,P_\ell^m]&=\frac{i}{2} \sqrt{\frac{2 \ell+1}{2 \ell-1}(\ell-m) (\ell-m-1)}\ P_{\ell-1}^{m+1}\\
&-\frac{i}{2} \sqrt{\frac{2 \ell+1}{2 \ell-1}(\ell+m) (\ell+m-1)}\ P_{\ell-1}^{m-1}\\
&+\frac{i}{2}  \frac{(\ell-1) (\ell+3)}{2 \ell+3} \sqrt{\frac{(\ell+m+1) (\ell+m+2)}{(2 \ell+1) (2 \ell+3)}}\ P_{\ell+1}^{m+1}\\
&-\frac{i}{2} \frac{(\ell-1) (\ell+3)}{2 \ell+3} \sqrt{\frac{ (\ell-m+1)(\ell-m+2)}{(2 \ell+1) (2 \ell+3)}}\ P_{\ell+1}^{m-1}\\
[K_3,P_\ell^m]&=i \sqrt{\frac{2 \ell+1}{2 \ell-1}(\ell-m) (\ell+m)}\ P_{\ell-1}^m,\\
&-i \frac{(\ell-1) (\ell+3)}{2 \ell+3} \sqrt{\frac{(\ell-m+1) (\ell+m+1)}{(2 \ell+1) (2 \ell+3)}}\ P_{\ell+1}^m.
\end{aligned}
\end{equation}

The form of the $\mathfrak{bms}_4$ algebra as given by Sachs~\cite{Sachs:1962zza} is recovered from the previous expression by multiplying the generators of super-translations by the factor
\begin{equation}
P^m_\ell\to \frac12\frac{(\ell+2)!}{(2\ell+1)!!} P^m_\ell
\end{equation}

One should note that Poincar\'e is a sub-algebra of $\mathfrak{bms}_4$, and no other finite-dimensional sub-algebras exist~\cite{Longhi:1997zt}. The Poincar\'e sub-algebra is spanned by $J_i,K_i,P_0^0,P_1^m$, with $i=1,2,3$ and $m=0,\pm1$. These operators form a closed sub-algebra of $\mathfrak{bms}_4$ because the matrix coefficients $(\mathcal J_i)_{\ell\, m'}^{\ell'm},(\mathcal K_i)_{\ell\, m'}^{\ell'm}$ vanish when $\ell=1$ and $\ell'\ge 2$.

We now proceed to perform the contraction of this algebra
in order to obtain its non-relativistic analogue.  Here and in the remainder of this document, we will place a hat over an object to indicate that it corresponds to the contracted, non-relativistic counterpart of the corresponding relativistic object. For example, the generators of non-relativistic super-translations will be denoted by $\hat P_\ell^m$, as opposed to the generators of relativistic super-translations, $P_\ell^m$.

As the Bargmann algebra contains a central charge, we expect that something similar happens in the case of $\mathfrak{nrbms}$. We therefore consider the direct product of the $\mathrm{BMS}$ group with $U(1)$, with generator $\tilde Z$, and introduce the following transformation 
\begin{equation}
\begin{split}
H&:=\omega(P^0_0+\tilde Z)\\
Z&:=\frac{1}{\omega}(P^0_0-\tilde Z)\\
\hat K_i&:=\frac{1}{\omega}K_i\\
\hat J_i&:= J_i
\end{split}\qquad\overset{\text{inverse}}{\Longleftrightarrow}\qquad
\begin{split}
P^0_0&=\frac{1}{2\omega}H+\omega Z\\
\tilde Z&=\frac{1}{2\omega}H-\omega Z\\
K_i&=\omega \hat K_i\\
J_i&=\hat J_i
\end{split}
\end{equation}
and
\begin{equation}
\hat P_\ell^m:=\omega^{f(\ell)}P_\ell^m\qquad\overset{\text{inverse}}{\Longleftrightarrow}\qquad P_\ell^m=\omega^{-f(\ell)}\hat P_\ell^m, \quad \ell\geq 1,
\end{equation}
where $f(\ell)$ is an unspecified function of $\ell$, and $\omega$ is a dimensionless parameter which we shall take $\omega\to\infty$ at the end.

In the limit $\omega\to\infty$, the centrally-extended relativistic algebra~\eqref{bmsalgebra} becomes
\begin{equation}\label{nrbmsalgebra}
\begin{split}
{}[\hat J_i,\hat J_j]&=i\epsilon_{ijk}\hat J_k\\
[\hat J_i,\hat K_j]&=i\epsilon_{ijk}\hat K_k\\
[\hat K_i,\hat K_j]&=0
\end{split}\hspace{40pt}
\begin{split}
[\hat P_\ell^m,H]&=0\\
[\hat J_i,H]&=0\\
[\hat K_i,H]&=i\hat P_i
\end{split}\hspace{40pt}
\begin{split}
[\hat P_\ell^m,\hat P_{\ell'}^{m'}]&=0\\
[\hat J_i,\hat P_\ell^m]&=i(\hat{\mathcal J}_i)_{\ell\, m'}^{\ell'm}\ \hat P_{\ell'}^{m'}\\
[\hat K_i,\hat P_\ell^m]&=i(\hat{\mathcal K}_i)_{\ell\, m'}^{\ell'm}\ \hat P_{\ell'}^{m'}
\end{split}
\end{equation}
where $(\hat{\mathcal J}_i)_{\ell\, m'}^{\ell'm}=(\mathcal J_i)_{\ell\, m'}^{\ell'm}$ and
\begin{equation}
(\hat{\mathcal K}_i)_{\ell\, m'}^{\ell'm}=\lim_{\omega\to\infty}\omega^{f(\ell)-f(\ell')-1}(\mathcal K_i)_{\ell\, m'}^{\ell'm}
\end{equation}

In order to have a non-trivial $\omega\to\infty$ limit, we must have
\begin{equation}
f(\ell)-f(\ell')-1\equiv0
\end{equation}
for some $\ell'$. Moreover, the structure constants $(\mathcal K_i)_{\ell\, m'}^{\ell'm}$ are non-zero only if $|\ell'-\ell|=1$. This means that the only non-trivial contractions are the ones that verify
\begin{equation}
f(\ell)=f(0)\pm\ell
\end{equation}

Furthermore, in order to obtain the Bargmann algebra as a sub-algebra,  for $\ell=1$ we should recover the standard contraction $\hat P_i=P_i$, so that $f(1)=0$. With this,
\begin{equation}\label{eq:two_signs}
f(\ell)=\pm(\ell-1)
\end{equation}

The conclusion of this discussion is that if we restrict ourselves to scalings of the form $\hat P_\ell^m=\omega^{f(\ell)}P_\ell^m$ then there are only two non-trivial contractions of the $\mathfrak{bms_4}$ algebra, corresponding to either sign in $f(\ell)=\pm(\ell-1)$, which we will call\footnote{One should note that this `$\pm$' has nothing to do with the conventional notation of $\mathrm{BMS}^\pm$ as the group that acts on the asymptotic past and future null infinity in asymptotically flat space-times.} $\mathfrak{nrbms}^\pm_4$. In either case, the structure constants are given by
\begin{equation}\label{nrmatrices}
\begin{aligned}
(\hat{\mathcal J}_i)_{\ell\, m'}^{\ell'm}&=(\mathcal J_i)_{\ell\, m'}^{\ell'm}\\
(\hat{\mathcal K}_i)_{\ell\, m'}^{\ell'm}&=\begin{cases}(\mathcal K_i)_{\ell\, m'}^{\ell'm} & \ell=\ell'\pm1\\ 0 & \text{otherwise}\end{cases}
\end{aligned}
\end{equation}

These two possible algebras contain time and space translations, rotations, Galilean boosts, spatial super-translations, and a central charge, and they both contain a Bargmann sub-algebra (see~Fig.\ref{fig:nrbms_algebra}). 
The matrices $\hat{\mathcal J}_i,\hat{\mathcal K}_i$ define a  non-unitary, reducible but indecomposable,  infinite-dimensional realisation of the homogeneous Galilei group, with algebra given by the first column of~\eqref{nrbmsalgebra}.

\begin{linespread}{1.0} \selectfont
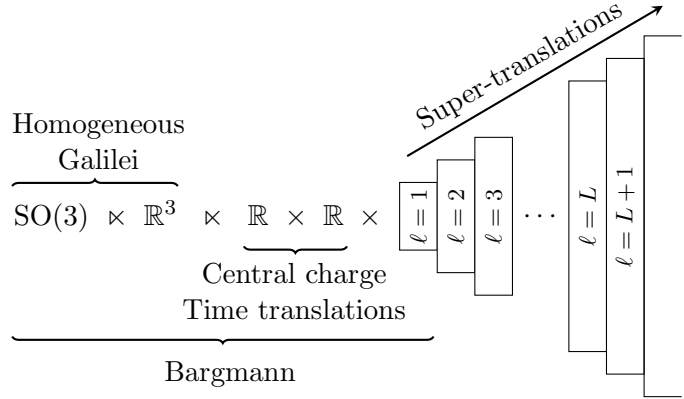
\begin{figure}[!h]
\centering
\begin{tikzpicture}
\node at (-1.45,0) {$\mathrm{SO}(3)\ \ltimes\ \mathbb R^3\quad\! \ltimes\,$};

\begin{scope}[shift={(1.05,0)}]
\filldraw[fill=white, draw=black] (1.2,-.45) rectangle (1.7,0.45) node[pos=.5, rotate=90] {\footnotesize$\ell=1$};
\filldraw[fill=white, draw=black] (1.7,-.75) rectangle (2.2,0.75) node[pos=.5, rotate=90] {\footnotesize$\ell=2$};
\filldraw[fill=white, draw=black] (2.2,-1.05) rectangle (2.7,1.05) node[pos=.5, rotate=90] {\footnotesize$\ell=3$};
\node at (3.1,0) {$\cdots$};
\filldraw[fill=white, draw=black] (3.45,-1.8) rectangle (3.95,1.8) node[pos=.5, rotate=90] {\footnotesize$\ell=L$};
\filldraw[fill=white, draw=black] (3.95,-2.1) rectangle (4.45,2.1) node[pos=.5, rotate=90] {\footnotesize$\ell=L+1$};
\draw[black] (4.95,-2.4) -- (4.45,-2.4) -- (4.45,2.4) -- (4.95,2.4);
\end{scope}

\draw[decoration={brace},decorate,thick]  (-2.9,.4) -- (-.7,.4);
\node at (-1.75,1) [align=center] {Homogeneous\\Galilei};
\node at (1.15,0) {$\mathbb R\ \times\ \mathbb R\ \,\times\,$};

\draw[decoration={brace,mirror},decorate,thick]  (.18,-.4) -- (1.55,-.4);

\node at (.85,-1) [align=center] {Central charge\\Time translations};

\draw[->,thick,>=stealth] (2.35,.8) -- (5.75,2.8);

\node at (3.85,2) [rotate=30] {Super-translations};

\draw[decoration={brace,mirror},decorate,thick]  (-2.9,-1.6) -- (2.7,-1.6);
\node at (0,-2.1) [align=center] {Bargmann};

\end{tikzpicture}
\caption{The structure of the $\mathfrak{nrbms}$ algebra. Each box represents an abelian algebra of dimension $2\ell+1$ corresponding to the super-translations $\{\hat P_\ell^m\}$, with $m=-\ell,\cdots,\ell$.}
\label{fig:nrbms_algebra}
\end{figure}
\end{linespread}

It is important to note that in the case of the $\mathfrak{nrbms}^+_4$ contraction, the boost operators lower the value of $\ell$ to $\ell-1$. This is in stark contrast with the relativistic case, where we have simultaneous contributions from $\ell-1$ and $\ell+1$. Therefore, unlike in the relativistic case, here the algebra resulting from the $+$ contraction contains an infinite number of finite-dimensional sub-algebras, obtained by considering all the super-translation generators with $1\le\ell\le L$ for a given $L$; the dimension of these sub-algebras is $8+\sum_{\ell=1}^{L}(2\ell+1)=L^2+2L+8$. All these sub-algebras contain a Bargmann sub-algebra, and the associated matrices
 $\hat{\mathcal J}_i$, $\hat{\mathcal K}_i$ provide an infinite number of finite-dimensional representations of the homogeneous Galilei group\footnote{Finite-dimensional indecomposable representations  
of homegeneous Galilei have been studied in \cite{Nikitin:2006mnn}.}, or Euclidean $E(3)$ group, of dimensions $\sum_{\ell=0}^{L}(2\ell+1)=(L+1)^2$, $L\geq 1$.

On the other hand, in the case of the $\mathfrak{nrbms}^-_4$ algebra, the boost operators raise the value of $\ell$ to $\ell+1$, which means that the only  finite-dimensional sub-algebra of $\mathfrak{nrbms}^-_4$ is the Bargmann algebra (see~Fig.\ref{fig:nrbms_algebra_pm}).

\begin{linespread}{1.0} \selectfont
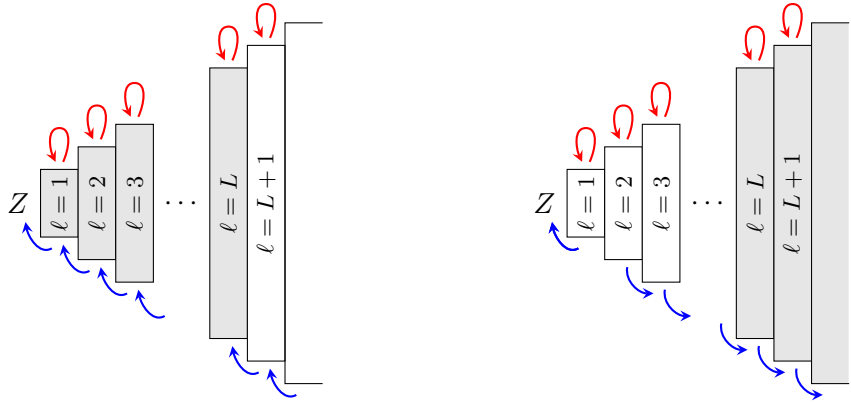
\begin{figure}[!h]
\centering
\begin{tikzpicture}

\node at (.9,0) {$Z$};

\draw [thick,red,->,>=stealth] (1.5,.55) to[out=65,in=0] (1.45,1) to[out=180,in=180-65] (1.4,.55);
\draw [thick,red,->,>=stealth] (1.5+.5,.55+.3) to[out=65,in=0] (1.45+.5,1+.3) to[out=180,in=180-65] (1.4+.5,.55+.3);
\draw [thick,red,->,>=stealth] (1.5+1,.55+.6) to[out=65,in=0] (1.45+1,1+.6) to[out=180,in=180-65] (1.4+1,.55+.6);
\draw [thick,red,->,>=stealth] (3.75,1.9) to[out=65,in=0] (3.7,2.35) to[out=180,in=180-65] (3.65,1.9);
\draw [thick,red,->,>=stealth] (3.75+.5,1.9+.3) to[out=65,in=0] (3.7+.5,2.35+.3) to[out=180,in=180-65] (3.65+.5,1.9+.3);

\filldraw[fill=black!10!white, draw=black] (1.2,-.45) rectangle (1.7,0.45) node[pos=.5, rotate=90] {\footnotesize$\ell=1$};
\filldraw[fill=black!10!white, draw=black] (1.7,-.75) rectangle (2.2,0.75) node[pos=.5, rotate=90] {\footnotesize$\ell=2$};
\filldraw[fill=black!10!white, draw=black] (2.2,-1.05) rectangle (2.7,1.05) node[pos=.5, rotate=90] {\footnotesize$\ell=3$};
\node at (3.1,0) {$\cdots$};
\filldraw[fill=black!10!white, draw=black] (3.45,-1.8) rectangle (3.95,1.8) node[pos=.5, rotate=90] {\footnotesize$\ell=L$};
\filldraw[fill=white, draw=black] (3.95,-2.1) rectangle (4.45,2.1) node[pos=.5, rotate=90] {\footnotesize$\ell=L+1$};
\draw[black] (4.95,-2.4) -- (4.45,-2.4) -- (4.45,2.4) -- (4.95,2.4);

\draw [thick,blue,->,>=stealth] (1.7+.15-.5,-.9+.3) to[out=180+30,in=-90+10] (1.2+.3-.5,-.55+.3);
\draw [thick,blue,->,>=stealth] (1.7+.15,-.9) to[out=180+30,in=-90+10] (1.2+.3,-.55);
\draw [thick,blue,->,>=stealth] (1.7+.15+.5,-.9-.3) to[out=180+30,in=-90+10] (1.2+.3+.5,-.55-.3);
\draw [thick,blue,->,>=stealth] (1.7+.15+1,-.9-.6) to[out=180+30,in=-90+10] (1.2+.3+1,-.55-.6);

\draw [thick,blue,->,>=stealth] (1.7+.15+2.25,-.9-1.35) to[out=180+30,in=-90+10] (1.2+.3+2.25,-.55-1.35);
\draw [thick,blue,->,>=stealth] (1.7+.15+2.25+.5,-.9-1.35-.3) to[out=180+30,in=-90+10] (1.2+.3+2.25+.5,-.55-1.35-.3);

\begin{scope}[shift={(7,0)}]
\node at (.9,0) {$Z$};

\draw [thick,red,->,>=stealth] (1.5,.55) to[out=65,in=0] (1.45,1) to[out=180,in=180-65] (1.4,.55);
\draw [thick,red,->,>=stealth] (1.5+.5,.55+.3) to[out=65,in=0] (1.45+.5,1+.3) to[out=180,in=180-65] (1.4+.5,.55+.3);
\draw [thick,red,->,>=stealth] (1.5+1,.55+.6) to[out=65,in=0] (1.45+1,1+.6) to[out=180,in=180-65] (1.4+1,.55+.6);
\draw [thick,red,->,>=stealth] (3.75,1.9) to[out=65,in=0] (3.7,2.35) to[out=180,in=180-65] (3.65,1.9);
\draw [thick,red,->,>=stealth] (3.75+.5,1.9+.3) to[out=65,in=0] (3.7+.5,2.35+.3) to[out=180,in=180-65] (3.65+.5,1.9+.3);

\filldraw[fill=white, draw=black] (1.2,-.45) rectangle (1.7,0.45) node[pos=.5, rotate=90] {\footnotesize$\ell=1$};
\filldraw[fill=white, draw=black] (1.7,-.75) rectangle (2.2,0.75) node[pos=.5, rotate=90] {\footnotesize$\ell=2$};
\filldraw[fill=white, draw=black] (2.2,-1.05) rectangle (2.7,1.05) node[pos=.5, rotate=90] {\footnotesize$\ell=3$};
\node at (3.1,0) {$\cdots$};
\filldraw[fill=black!10!white, draw=black] (3.45,-1.8) rectangle (3.95,1.8) node[pos=.5, rotate=90] {\footnotesize$\ell=L$};
\filldraw[fill=black!10!white, draw=black] (3.95,-2.1) rectangle (4.45,2.1) node[pos=.5, rotate=90] {\footnotesize$\ell=L+1$};
\filldraw[fill=black!10!white,draw=white] (4.45,-2.4) rectangle (4.95,2.4);
\draw[black] (4.95,-2.4) -- (4.45,-2.4) -- (4.45,2.4) -- (4.95,2.4);

\draw [thick,blue,->,>=stealth] (1.7+.15-.5,-.9+.3) to[out=180+30,in=-90+10] (1.2+.3-.5,-.55+.3);

\draw [thick,blue,->,>=stealth] (1.7+.15-.5,-.9+.3) to[out=180+30,in=-90+10] (1.2+.3-.5,-.55+.3);
\draw [thick,blue,<-,>=stealth] (1.7+.15+1,-.9-.6) to[out=180,in=-90-10] (1.2+.3+1,-.55-.6);
\draw [thick,blue,<-,>=stealth] (1.7+.15+.5,-.9-.3) to[out=180,in=-90-10] (1.2+.3+.5,-.55-.3);
\draw [thick,blue,<-,>=stealth] (1.7+.15+2.25-.5,-.9-1.35+.3) to[out=180,in=-90-10] (1.2+.3+2.25-.5,-.55-1.35+.3);
\draw [thick,blue,<-,>=stealth] (1.7+.15+2.25,-.9-1.35) to[out=180,in=-90-10] (1.2+.3+2.25,-.55-1.35);
\draw [thick,blue,<-,>=stealth] (1.7+.15+2.25+.5,-.9-1.35-.3) to[out=180,in=-90-10] (1.2+.3+2.25+.5,-.55-1.35-.3);

\end{scope}

\end{tikzpicture}
\caption{The inhomogeneous part of $\mathfrak{nrbms}^+$ (left) and $\mathfrak{nrbms}^-$ (right) . The upper row  arrows represent the action of rotations  on the generators of super-translations, while the lower row ones represent that of the Galilean boosts. Rotations do not change the value of $\ell$, while boosts take us from $\ell$ to $\ell\mp1$. The grey boxes represent the different sub-algebras, obtained from varying $L$ (in the first case, they are all finite-dimensional, while in the second case they are all infinite-dimensional).}
\label{fig:nrbms_algebra_pm}
\end{figure}
\end{linespread}


In the following section we will construct an explicit realisation of the $\mathfrak{nrbms}^+_4$ algebra corresponding to the plus sign contraction, and we will also discuss the possibility of adding dilatations and expansions in order to obtain a {Schr\"odinger}-$\mathfrak{nrbms}$.

\section{A canonical realisation of the $\mathfrak{nrbms}_4$ algebra} 
\label{section3}

We now proceed to construct  a canonical realisation of the $\mathfrak{nrbms}^+_4$ algebra.   Our construction will be analogous to the canonical construction of the $\mathfrak{bms}_4$ algebra, as given by Longhi and Materassi \cite{Longhi:1997zt} (see also~\cite{Gomis:2015ata},~\cite{Batlle:2017llu}). The starting point is a free complex Schr\"odinger field $\psi(t, {\boldsymbol x})$ in $1+3$ Galilean space-time, with action
\begin{equation}
S[\psi]=\int\mathrm dt\,\mathrm d\boldsymbol x\, \left(i\psi^*(t, \boldsymbol x)\dot\psi(t, \boldsymbol x)+\frac{1}{2\mu}\psi^*(t, \boldsymbol x)\nabla^2\psi(t,\boldsymbol x)\right),
\end{equation}
whose equation of motion is the Schr\"odinger equation,
\begin{equation}
i\dot\psi+\frac{1}{2\mu}\nabla^2\psi=0
\end{equation}
The action
is invariant under the general Bargmann transformation
\begin{equation}
\begin{aligned}
\delta\psi(t, \boldsymbol x) = &-i\mu \eta \psi(t, \boldsymbol x) + \epsilon \dot{\psi}(t, \boldsymbol x) - a_i \partial_i\psi(t, \boldsymbol x) - \beta_i (t \partial_i - i\mu x_i )\psi(t, \boldsymbol x)\\
& - \omega_i \epsilon_{ijk} (x_j \partial_k -x_k \partial_j)\psi(t, \boldsymbol x),
\end{aligned}
\end{equation}
where $\eta$, $\epsilon$, $a_i$, $\beta_i$ and $\omega_i$ are, respectively, parameters corresponding to the central charge, time translations, spatial translations, Galilean boosts and spatial rotations. The associated Noether charges are
\begin{equation}
\begin{aligned}
Z&=\mu\int\mathrm d\boldsymbol x\ |\psi|^2\\
H&=\frac{1}{2\mu}\int\mathrm d\boldsymbol x\ |\nabla\psi|^2\\
\hat P_i&=-i\int\mathrm d\boldsymbol x\ \psi^*\partial_i\psi,\\
\hat K_i&=t\hat P_i-\mu\int\mathrm d \boldsymbol x\ \psi^*x_i\psi\\
\hat J_i&=-i\epsilon_{ijk}\int\mathrm d \boldsymbol x\ \psi^*x_j\partial_k\psi,
\end{aligned}
\end{equation}
and they satisfy the Bargmann algebra,
\begin{equation}
\begin{split}
\{\hat J_i, \hat J_j\}&=i\epsilon_{ijk}\hat J_k\\
\{\hat J_i,\hat K_j\}&=i\epsilon_{ijk}\hat K_k\\
\{\hat K_i,\hat K_j\}&=0
\end{split}\hspace{60pt}
\begin{split}
\{\hat  J_i,\hat P_j\}&=i\epsilon_{ijk} \hat P_k\\
\{\hat K_i, \hat P_j\}&=i Z\delta_{ij}\\
\{\hat K_i,H\}&=i\hat  P_i,
\end{split}
\end{equation}
where $\{\cdot,\cdot\}$ denotes the Dirac bracket (such that the canonical algebra reads $\{\psi(t,\boldsymbol x),\psi^*(t,\boldsymbol y)\}=-i\delta(\boldsymbol x-\boldsymbol y)$).

The field $\psi(t,\boldsymbol x)$, when on-shell, can be expanded in Fourier modes as
\begin{equation}\label{eq:sch_sol}
\psi(t,\boldsymbol{x})=\int\frac{\mathrm d\boldsymbol k}{(2\pi)^3}\ \mathrm e^{-ik^0 t+ i \boldsymbol{x}\cdot\boldsymbol{k}}a(\boldsymbol k),
\end{equation}
where $k^0=\frac{1}{2\mu}\boldsymbol k^2$.
 The absence of $a^*$ in the expansion of $\psi$ reflects the fact that no antiparticles would be present in the corresponding second quantised theory.


When expressed in terms of $a,a^*$, the on-shell Noether charges become
\begin{equation}\label{eq:non-relativistic_noether}
\begin{aligned}
Z&=\mu\int\frac{\mathrm d\boldsymbol k}{(2\pi)^3}\ a^*(\boldsymbol k) a(\boldsymbol k)\\
H&=\int\frac{\mathrm d\boldsymbol k}{(2\pi)^3}\ a^*(\boldsymbol k)\frac{\boldsymbol k^2}{2\mu} a(\boldsymbol k)\\
\hat P_i&=\int\frac{\mathrm d\boldsymbol k}{(2\pi)^3}\ a^*(\boldsymbol k)k_i a(\boldsymbol k)\\
\hat K_i&=t\hat P_i-i\mu\int\frac{\mathrm d\boldsymbol k}{(2\pi)^3}\ a^*(\boldsymbol k)\partial_i a(\boldsymbol k)\\
\hat J_i&=-i\epsilon_{ijk}\int\frac{\mathrm d\boldsymbol k}{(2\pi)^3}\ a^*(\boldsymbol k)k_j\partial_k a(\boldsymbol k).
\end{aligned}
\end{equation}

As far as the homogeneous Galilei algebra is concerned, the time dependent part $t\hat P_i$ in $\hat K_i$ is irrelevant, and so we will drop it in what follows. Using $i\{a(\boldsymbol k),a^*(\boldsymbol k')\}=(2\pi)^3\delta(\boldsymbol k-\boldsymbol k')$, one may easily check that
\begin{equation}
\begin{aligned}
i\{\hat K_i,a(\boldsymbol k)\}&=\hat{\mathcal K}_ia(\boldsymbol k)\\
i\{\hat J_i,a(\boldsymbol k)\}&=\hat{\mathcal J}_ia(\boldsymbol k),
\end{aligned}
\end{equation}
where $\hat{\mathcal K}_i:=i\mu\partial_i$ and $\hat{\mathcal J}_i:=i\epsilon_{ijk}k_j\partial_k$ are differential operators in $\boldsymbol{k}$ space, which provide  a realisation of the homogeneous Galilei algebra,
\begin{equation}
\begin{aligned}
{}[\hat{\mathcal J}_i, \hat{\mathcal J}_j]&=-i\epsilon_{ijk}\hat{\mathcal J}_k\\
[\hat{\mathcal J}_i,\hat{\mathcal K}_j]&=-i\epsilon_{ijk}\hat{\mathcal K}_k\\
[\hat{\mathcal K}_i,\hat{\mathcal K}_j]&=0,
\end{aligned}
\end{equation}
where $[\cdot,\cdot]$ denotes a commutator.

We now consider the 
quadratic Casimir operator of the homogeneous Galilei group, 
$\hat \Delta:=\hat{\mathcal K}_i^2$, that is,
\begin{equation}
\hat \Delta=-\mu^2\partial_i^2
\end{equation}

The momenta $k_i$ satisfy an eigenvalue equation with respect to $\hat \Delta$,
\begin{equation}
\hat \Delta k_i=0,
\end{equation}
i.e., the momenta are zero-modes of $\hat \Delta$. As in the relativistic construction \cite{Longhi:1997zt}, this motivates us to consider the most general zero-mode equation
\begin{equation}
\hat \Delta\,\hat \chi(\boldsymbol k)=0,
\end{equation}
whose set of solutions $\{\hat\chi\}$ will contain the momenta $k_i$ as a subset. The general set zero-modes will correspond to the complete set of super-translations, in the same sense that $k_i$ corresponds to standard translations (cf.~\eqref{eq:non-relativistic_noether}). It bears mentioning that the second Casimir, $\hat{\mathcal J}_i\hat{\mathcal K}_i$, vanishes identically.

The general solution of the equation above is
\begin{equation}
\hat \chi_\ell^m(r,\theta,\varphi)=f_\ell(r)Y_\ell^m(\theta,\varphi)
\end{equation}
where $r,\theta,\varphi$ are the spherical coordinates of the dimensionless momentum $\frac{1}{\mu}\boldsymbol k$, and $Y_\ell^m$ are the spherical harmonics, with $\ell,m$ integers such that $|m|\le \ell$. Here, $f_\ell$ is given by the solution of
\begin{equation}
r^2f''_\ell+2rf'_\ell-\ell(\ell+1)f_\ell=0
\end{equation}
that is,
\begin{equation}
f_\ell(r)=c_1r^\ell+c_2r^{-(\ell+1)}
\end{equation}

We are looking for zero-modes $\hat\chi$ that are regular at the origin, like the three momenta $k_i$, so we drop the second solution. With this,
\begin{equation}\label{eq:non-relativistic_omega}
\hat \chi_\ell^m(r,\theta,\varphi)=r^\ell Y_\ell^m(\theta,\varphi)
\end{equation}

It is easy to see that the modes corresponding to $\ell=1$ agree with the spherical components of $\frac{1}{\mu}\boldsymbol k$, so that, as expected, the family $\{\hat \chi_\ell^m\}$ contains the functions $k_i$ as a special subcase.

As the angular part of the modes $\hat \chi_\ell^m$ is given by the spherical harmonics, these functions satisfy
\begin{equation}\label{eq:D_chi_algebra}
\begin{aligned}
\hat{\mathcal J}_1\hat \chi_\ell^m&=\frac{i}{2} \sqrt{(\ell+m) (\ell-m+1)}\ \hat \chi_\ell^{m-1}-\frac{i}{2}\sqrt{(\ell-m) (\ell+m+1)}\ \hat \chi_\ell^{m+1}\\
\hat{\mathcal J}_2\hat \chi_\ell^m&=\frac{1}{2} \sqrt{(\ell+m) (\ell-m+1)}\ \hat \chi^{m-1}_\ell+\frac{1}{2} \sqrt{(\ell-m) (\ell+m+1)}\ \hat \chi^{m+1}_\ell\\
\hat{\mathcal J}_3\hat \chi_\ell^m&=m\, \hat \chi^m_\ell
\end{aligned}
\end{equation}

Moreover, 
the boost differential operators $\hat{\mathcal K}_i$ do also have a simple action on the modes $\hat \chi_\ell^m$:
\begin{equation}\label{eq:C_chi_algebra}
\begin{aligned}
\hat{\mathcal K}_1\hat \chi_\ell^m&=\frac{1}{2} \sqrt{\frac{2 \ell+1}{2 \ell-1} (\ell-m) (\ell-m-1)}\ \hat \chi_{\ell-1}^{m+1}+\frac{1}{2} \sqrt{\frac{2 \ell+1}{2 \ell-1}(\ell+m) (\ell+m-1)}\ \hat \chi_{\ell-1}^{m-1}\\
\hat{\mathcal K}_2\hat \chi_\ell^m&=\frac{i}{2} \sqrt{\frac{2 \ell+1}{2 \ell-1}(\ell-m) (\ell-m-1)}\ \hat \chi_{\ell-1}^{m+1}-\frac{i}{2} \sqrt{\frac{2 \ell+1}{2 \ell-1}(\ell+m) (\ell+m-1)}\ \hat \chi_{\ell-1}^{m-1}\\
\hat{\mathcal K}_3\hat \chi_\ell^m&=i \sqrt{\frac{2 \ell+1}{2 \ell-1}(\ell-m) (\ell+m)}\ \hat \chi_{\ell-1}^m
\end{aligned}
\end{equation}

With this, we now define the generators of super-translations  as
\begin{equation}\label{nrPl}
\hat P_\ell^m:= \mu\int\frac{\mathrm d\boldsymbol k}{(2\pi)^3}\ a^*(\boldsymbol k)\hat \chi_\ell^m(\boldsymbol k) a(\boldsymbol k).
\end{equation}
Unlike the relativistic case~\cite{Longhi:1997zt}, here the existence of $\hat P_\ell^m$ is not guaranteed by the existence of $\hat P_i$, because the non-relativistic modes $\hat \chi$ scale as $r^\ell$ for large $r$ instead of linearly with $r$. Therefore, $a(\boldsymbol k)$ being square-integrable is not enough for the integral defining $\hat P_\ell^m$ to converge; we must impose the stronger condition that $|\boldsymbol k|^\ell |a(\boldsymbol k)|^2$ is integrable for all $\ell\in\mathbb N$. We also note that, due to the form of $\hat P_\ell^m$ in~\eqref{nrPl}, the quantum theory would yield an unbroken realisation of the symmetry, i.e., the Fock vacuum would be invariant.

Using~\eqref{eq:D_chi_algebra} and~\eqref{eq:C_chi_algebra}, we see that the functions $\hat P_\ell^m$ satisfy the algebra
\begin{equation}
\begin{aligned}
\{\hat J_1,\hat P_\ell^m\}&=\frac{i}{2} \sqrt{(\ell+m) (\ell-m+1)}\ \hat P_\ell^{m-1}-\frac{i}{2}\sqrt{(\ell-m) (\ell+m+1)}\ \hat P_\ell^{m+1}\\
\{\hat J_2,\hat P_\ell^m\}&=\frac{1}{2} \sqrt{(\ell+m) (\ell-m+1)}\ \hat P^{m-1}_\ell+\frac{1}{2} \sqrt{(\ell-m) (\ell+m+1)}\ \hat P^{m+1}_\ell\\
\{\hat J_3,\hat P_\ell^m\}&=m\, P^m_\ell
\end{aligned}
\end{equation}
and
\begin{equation}
\begin{aligned}
\{\hat K_1,\hat P_\ell^m\}&=\frac{1}{2} \sqrt{\frac{2 \ell+1}{2 \ell-1}(\ell-m) (\ell-m-1)}\ \hat P_{\ell-1}^{m+1}\\
&+\frac{1}{2} \sqrt{\frac{2 \ell+1}{2 \ell-1}(\ell+m) (\ell+m-1)}\ \hat P_{\ell-1}^{m-1}\\
\{\hat K_2,\hat P_\ell^m\}&=\frac{i}{2} \sqrt{\frac{2 \ell+1}{2 \ell-1}(\ell-m) (\ell-m-1)}\ \hat P_{\ell-1}^{m+1}\\
&-\frac{i}{2} \sqrt{\frac{2 \ell+1}{2 \ell-1} (\ell+m) (\ell+m-1)}\ \hat P_{\ell-1}^{m-1}\\
\{\hat K_3,\hat P_\ell^m\}&=i \sqrt{\frac{2 \ell+1 }{2 \ell-1}(\ell-m) (\ell+m)}\ \hat P_{\ell-1}^m
\end{aligned}
\end{equation}
which constitutes a realisation of the $\mathfrak{nrbms}^+_4$ algebra as given by \eqref{nrbmsalgebra},~\eqref{nrmatrices}. 
 In principle, one may construct a realisation of $\mathfrak{nrbms}_4^-$ by using the second solution of the radial equation, $f_\ell\sim r^{-(\ell+1)}$, but the fact that these functions are singular at the origin implies that the Fourier modes $a(\boldsymbol k)$ must go to zero faster than any polynomial if we want the integral that defines $\hat P_\ell^m$ to converge. We will not consider this possibility any further here.

{ This constitutes our proposal for the non-relativistic $\mathfrak{bms}$ algebra. It is important to stress that we obtained the same algebra both by means of the abstract group contraction, and the explicit canonical realisation. This agreement is in fact not completely unexpected: one may check that the explicit realisation is nothing but the $m\to\infty$ limit of the canonical realisation of the relativistic $\mathfrak{bms}$ algebra, as given in~\cite{Longhi:1997zt}.}

Unlike in the relativistic case,  the modes $\hat\chi(\boldsymbol k)$ are actually homogeneous polynomials in $k_i$, of degree $\ell$. This means that the symmetries generated by $\hat P_\ell^m$ are local when acting on the field $\psi$, meaning that 
\begin{equation}
\delta_\ell^m\psi=\{\hat{P}_\ell^m,\psi\}=\hat\chi_\ell^m(-i\partial)\psi
\end{equation}
where $\hat\chi_\ell^m$ is a harmonic polynomial of degree $\ell$. In other words, $\hat \chi_\ell^m(-i\partial)$ is nothing but a polynomial in $\partial_i$:
\begin{equation}
\hat \chi_\ell^m(-i\partial)=\sum_{|\alpha|=\ell}c^\alpha\partial_\alpha
\end{equation}
for a certain set of coefficients $c_\alpha$, and where $\alpha$ is a multi-index. Differential operators of this form (and generalisations thereof), in the context of symmetries of partial differential equations, have been studied extensively in the literature; see for example \cite{Nikitin:1991}\cite{Valenzuela:2009gu}.  

{ Needless to say, for the operation $\delta_\ell^m\psi=\hat\chi_\ell^m(-i\partial)\psi$ to be symmetry of the system, it must be well-defined: $\psi$ must have at least $\ell$ (spatial) derivatives, and all of them must be square integrable (so that they represent valid wave-functions). In other words, we must have $\psi(t,\cdot)\in W^{\ell,2}(\mathbb R^3,\mathbb C,\mathrm d\boldsymbol x)$, the Sobolev space of order $\ell$. One should note that if $\psi(t,\cdot)\in W^{\ell,2}(\mathbb R^3,\mathbb C,\mathrm d\boldsymbol x)$ then $|\boldsymbol k|^\ell a\in L^2(\mathbb R^3,\mathbb C,\mathrm d\boldsymbol k)$, and therefore the integral that defines $P_\ell^m$, to wit~\eqref{nrPl}, converges, as one would expect.}

The operators $P_\ell^m$ are symmetries of the Schr\"odinger differential operator $\partial_t-\{H,\cdot\}$. Indeed, if $\psi(x)$ is a solution of the Schr\"odinger equation,
then so is $\psi(x)+\delta_\ell^m\psi(x)$ for any $\ell,m$. To see this, we note that the  general solution of the Schr\"odinger equation is of the 
form~\eqref{eq:sch_sol}
\begin{equation}\label{eq:sch_sol_2}
\psi(x)=\int\frac{\mathrm d\boldsymbol k}{(2\pi)^3}\ \mathrm e^{-ik^0 t + i \boldsymbol{k}\cdot\boldsymbol{x}}a(\boldsymbol k)
\end{equation}
and therefore
\begin{equation}
\delta_\ell^m\psi(x)=\int\frac{\mathrm d\boldsymbol k}{(2\pi)^3}\ \mathrm e^{-ik^0 t + i \boldsymbol{k}\cdot\boldsymbol{x}}\hat\chi_\ell^m(\boldsymbol k)a(\boldsymbol k)
\end{equation}
which is itself of the form~\eqref{eq:sch_sol_2}, with $a(\boldsymbol k)\to \hat\chi_\ell^m(\boldsymbol k)a(\boldsymbol k)$.

Furthermore, the fact that the polynomials $\hat\chi_\ell^m(\boldsymbol k)$ are \emph{homogeneous} and of degree $\ell$ implies that they satisfy
\begin{equation}
\mathcal D\hat\chi_\ell^m=\ell\hat\chi_\ell^m
\end{equation}
where $\mathcal D$ is the homogeneity operator, $\mathcal D:=k^i\partial_i$. Hence, if we define the dilatation operator as
\begin{equation}
D:=2tH+i\int\frac{\mathrm d\boldsymbol k}{(2\pi)^3}\ a^*(\boldsymbol k) \mathcal D a(\boldsymbol k)
\end{equation}
then the generators of super-translations satisfy
\begin{equation}
\{D,\hat P_\ell^m\}=i\ell \hat P_\ell^m
\end{equation}
which extends the $\mathfrak{nrbms}^+_4$ algebra to include dilatations (which are also symmetries of the Schr\"odinger action), giving rise to a Weyl-$\mathfrak{nrbms}$.

Once we check that the algebra admits dilatations, it becomes natural to ask ourselves about the action of the expansion operator (or Schr\"odinger conformal transformations), given by
\begin{equation}\label{eq:conformal}
C:=-t^2H+tD+\frac\mu2\int\frac{\mathrm d\boldsymbol k}{(2\pi)^3}\ a^*(\boldsymbol k) \hat \Delta a(\boldsymbol k).
\end{equation}

Using
\begin{equation}
\begin{aligned}
{}[\hat \Delta,\hat \chi_\ell^m]&=\hat\Delta\hat \chi_\ell^m+2(\partial_i\hat \chi_\ell^m)\frac{\partial}{\partial k^i}\\
&=\frac2\mu(\hat{\mathcal K}_i)_{\ell\, m'}^{\ell'm}\ \hat \chi_{\ell'}^{m'}\frac{\partial}{\partial k^i}
\end{aligned}
\end{equation}
we obtain
\begin{equation}\label{expansionalgebra}
\{C,\hat P_\ell^m\}= it\ell \hat P_\ell^m+\mu\int\frac{\mathrm d\boldsymbol k}{(2\pi)^3}\ a^*(\boldsymbol k) \left[(\hat{\mathcal K}_i)_{\ell\, m'}^{\ell'm}\ \hat\chi_{\ell'}^{m'}(\boldsymbol k)\frac{\partial}{\partial k^i}\right] a(\boldsymbol k)
\end{equation}

The \emph{r.h.s.} of~\eqref{expansionalgebra} is not an element of $\mathfrak{nrbms}^+_4$. This means that if we attempt to extend the algebra to include $C$, the resulting algebra is not closed, which seems to preclude a possible Schr\"odinger-$\mathfrak{nrbms}$ with only super-translations (that is, without further extending the set of generators).

In any case, the \emph{r.h.s.}~\eqref{expansionalgebra} is the bracket of two conserved quantities, which means that it is conserved as well. Indeed,
\begin{equation}
\left[\frac{\partial}{\partial t}+\{\cdot,H\}\right]\{C,\hat P_\ell^m\}=\int\frac{\mathrm d\boldsymbol k}{(2\pi)^3}\ a^*(\boldsymbol k)\left[k^i(\hat{\mathcal K}_i)_{\ell\, m'}^{\ell'm}\ \hat\chi_{\ell'}^{m'}(\boldsymbol k)-i\ell\hat\chi_{\ell}^{m}(\boldsymbol k) \right]a(\boldsymbol k)
\end{equation}

Using $i\mu\partial_i\hat\chi_\ell^m=i(\hat{\mathcal K}_i)_{\ell\, m'}^{\ell'm}\ \hat\chi_{\ell'}^{m'}$ it follows that the factor in brackets vanishes, which means that $\{C,\hat P_\ell^m\}$, despite not being an element of $\mathfrak{nrbms}$, is a conserved operator i.e., it generates a symmetry of the Schr\"odinger equation.  In the particular case $\ell=1$, this commutator agrees with the generators of boosts, $\{C,\hat P_i\}=i\hat K_i$. It is tempting to let $\{C,\hat P_\ell^m\}$ define a new kind of generator of symmetries, which would generalise the standard generators of boosts; we could dub these objects \emph{super-boosts}. It will be interesting to explore their relation to the relativistic super-rotations.

\section{Conclusions and Outlook}\label{section4}

We have found the two unique contractions of $\mathfrak{bms}_4\times \mathfrak{u}(1)$ that contain as a sub-algebra the Bargmann algebra. 
The $\mathfrak{nrbms}^+_4$ algebra contains  an infinite number of finite-dimensional sub-algebras, obtained by considering the first $L\in\mathbb N$ super-translations $\hat P_\ell^m$. We have also found an infinite number of finite-dimensional indecomposable representations of the homogeneous Galilei algebra given by the matrices $\hat{\mathcal J}_i,\hat{\mathcal K}_i$.  The Bargmann algebra 
is the only finite dimensional sub-algebra of $\mathfrak{nrbms}^-_4$, the associated indecomposable representation of the homogeneous Galilei group being infinite-dimensional.

We found a canonical realisation of $\mathfrak{nrbms}^+_4$  in terms of the 
Fourier modes of a free  Schr\"odinger field. We have seen that we can add a dilation generator to $\mathfrak{nrbms}^+_4$, however we cannot extend the algebra with the expansion that is present in the Schr\"odinger algebra. 

One may presume that the conformal symmetry can only be present in the case of an extended $\mathrm{NRBMS}$, that is, in the group that includes 
super-boosts, as defined by the right-hand side of \eqref{expansionalgebra}, and
super-rotations. 

The extension of the previous results to other dimensions is under study~\cite{future}. In the case of $1+2$ dimensions we expect  to obtain an exotic $\mathfrak{nrbms}$ algebra with two central charges. 

It will be interesting to study the asymptotic symmetries of asymptotically flat Newtonian space-times and to check whether or not the algebra coincides with the one of the two $\mathfrak{nrbms}$ algebras we have constructed.

\section*{Acknowledgements}\addcontentsline{toc}{section}{Acknowledgements}
We acknowledge interesting discussions with   Eric Bergshoeff, Roberto Casalbuoni and Axel Kleinschmidt.
JG has been supported  by FPA2013-46570-C2-1-P, 2014-SGR-104 (Generalitat de Cata\-lunya) and Consolider CPAN and by
the Spanish goverment (MINECO/FEDER) under project MDM-2014-0369 of ICCUB (Unidad de Excelencia Mar\'\i a de Maeztu).
CB is partially supported by  the Generalitat de Catalunya through project 2014 SGR 267 and by the Spanish government (MINECO/FEDER) under project CICYT DPI2015-69286-C3-2-R.




\begin{thebibliography}{99}\addcontentsline{toc}{section}{References}

\bibitem{BMS-1}
H.~Bondi, MGJ.~Van der Burg and AWK.~Metzner,
``Gravitational waves in general relativity. VII. Waves from axi-symmetric isolated systems'',
Proc. Roy. Soc. Lond. A 269, 21 (1962); R. K. Sachs, 
``Gravitational waves in general relativity VIII. Waves in asymptotically flat space-time'',
Proc. Roy. Soc. Lond. A 270, 103 (1962).


\bibitem{Strominger:2017} 
A.~Strominger,
``Lectures on the Infrared Structure of Gravity and Gauge Theory,''
\href{http://arxiv.org/abs/arXiv:1703.05448}{arXiv:1703.05448 [hep-th]}.

\bibitem{Banks:2003vp} 
  T.~Banks,
  ``A Critique of pure string theory: Heterodox opinions of diverse dimensions,''
  \href{http://arxiv.org/abs/hep-th/0310288}{hep-th/0306074}.


\bibitem{deBoer:2003vf} 
  J.~de Boer and S.~N.~Solodukhin,
  ``A Holographic reduction of Minkowski space-time,''
  Nucl.\ Phys.\ B {\bf 665}, 545 (2003).
  \href{http://arxiv.org/abs/hep-th/0303006}{hep-th/0303006}.


\bibitem{Arcioni:2003xx} 
  G.~Arcioni and C.~Dappiaggi,
  ``Exploring the holographic principle in asymptotically flat space-times via the BMS group,''
  Nucl.\ Phys.\ B {\bf 674}, 553 (2003).
  \href{http://arxiv.org/abs/hep-th/0306142}{hep-th/0306142}.


\bibitem{Barnich:2010eb} 
  G.~Barnich and C.~Troessaert,
  ``Aspects of the BMS/CFT correspondence,''
  JHEP {\bf 1005}, 062 (2010).
  \href{http://arxiv.org/abs/arXiv:1001.1541}{arXiv:1001.1541 [hep-th]}.
  
\bibitem{Duval:2014}
C.~Duval, G.~W.~Gibbons and P.~A.~Horvathy, ``Conformal Carroll groups,'' \href{http://arxiv.org/abs/1403.4213}{{\tt arXiv:1403.4213 [hep-th]}}; C.~Duval, G.~W.~Gibbons and P.~A.~Horvathy, ``Conformal Carroll groups and BMS symmetry,'' \href{http://arxiv.org/abs/1402.5894}{{\tt arXiv:1402.5894 [gr-qc]}}.




\bibitem{Levy-Leblond}
J.M.~L\'evy-Leblond, ``Une nouvelle limite non-relativiste du group de Poincar\'e'', {\em Ann.~Inst.~H.~Poincar\'e} {\bf 3} (1965) 1; V. D. Sen Gupta, ``On an Analogue of the Galileo Group,'' {\em Il Nuovo Cimento} {\bf 54} (1966) 512.

\bibitem{Sachdev}
S.~Sachdev,
``Quantum phase transitions,''
Cambridge University Press (2011).
ISBN 978-0-521-51468-2.



\bibitem{liu}
Y.~Liu, K.~Schalm, W.~Sun, J. ~Zaanen.
``Holographic duality in condensed matter  physics,''
Cambridge University Press (2015).
ISBN 9781107080089.

\bibitem{Hartnoll:2016apf}
  S.~A.~Hartnoll, A.~Lucas and S.~Sachdev,
  ``Holographic quantum matter,''
  arXiv:1612.07324 [hep-th].

\bibitem{Son:2013rqa}
  D.~T.~Son,
  ``Newton-Cartan Geometry and the Quantum Hall Effect,''
  arXiv:1306.0638 [cond-mat.mes-hall].

\bibitem{Geracie:2016bkg}
  M.~Geracie,
  ``Galilean Geometry in Condensed Matter Systems,''
  arXiv:1611.01198 [hep-th].


\bibitem{future} C.~Batlle, D.~Delmastro and J.~Gomis (work in progress).

 \bibitem{Longhi:1997zt}
 G.~Longhi and M.~Materassi,
 ``A Canonical realization of the BMS algebra,''
 J.\ Math.\ Phys.\  {\bf 40} (1999) 480.
 \href{http://arxiv.org/abs/hep-th/9803128}{arXiv:9803128 [hep-th]}.

\bibitem{Sachs:1962zza}
R. K. Sachs ``Asymptotic symmetries in gravitational theory"
Phys. Rev. 128, 2851 (1962).


\bibitem{Cartan:1923zea}
  E.~Cartan,
  ``Sur les vari\'et\'es \`a connexion affine et la th\'eorie de la relativit\'e g\'en\'eralis\'ee. (premi\`ere partie),''
  Annales Sci.\ Ecole Norm.\ Sup.\  {\bf 40} (1923) 325.

\bibitem{Duval:1984cj}
  C.~Duval, G.~Burdet, H.~P.~Kunzle and M.~Perrin,
  Phys.\ Rev.\ D {\bf 31} (1985) 1841.
  doi:10.1103/PhysRevD.31.1841

\bibitem{Bargmann:1954gh}
  V.~Bargmann,
 ``On Unitary ray representations of continuous groups,''
  Annals Math.\  {\bf 59} (1954) 1.
  doi:10.2307/1969831


\bibitem{Barnich:2011ct} 
G.~Barnich and C.~Troessaert,
``Supertranslations call for superrotations,''
PoS, 010 (2010),
Ann.\ U.\ Craiova Phys.\  {\bf 21}, S11 (2011).
\href{http://arxiv.org/abs/1102.4632}{\tt arXiv:1102.4632 [gr-qc]}.

\bibitem{Nikitin:2006mnn}
M.~de~Montigny, J.~Niederle and A.~G.~Nikitin,
``Galilei invariant theories. I. Constructions of indecomposable finite-dimensional representations of the homogeneous Galilei group: directly and via contractions,"
J. Phys. A: Math. and Theor. \textbf{39}(29), pp. 9365-9385 (2006).
\href{https://arxiv.org/abs/math-ph/0604002}{arXiv:0604002 [math-th]}.



\bibitem{Gomis:2015ata}
J.~Gomis and G.~Longhi,
``Canonical realization of Bondi-Metzner-Sachs symmetry: Quadratic Casimir,''
Phys.\ Rev.\ D {\bf 93} (2016) no.2,  025030.
doi:10.1103/PhysRevD.93.025030
\href{http://arxiv.org/abs/arXiv:1508.00544}{arXiv:1508.00544 [hep-th]}.

\bibitem{Batlle:2017llu} 
C.~Batlle, V.~Campello and J.~Gomis,
``Canonical Realization of BMS$_3$,''
\href{http://arxiv.org/abs/arXiv:1703.01833}{arXiv:1703.01833 [hep-th]}.

\bibitem{Nikitin:1991}
A.~G.~Nikitin, 
``Complete set of symmetry operators of the Schrödinger equation,"
Ukrainian Mathematical Journal, vol. 43 (1991) no. 11, pp. 1413-1418. [Translated from Ukrainskii Matematicheskii Zhurnal, vol. 43 (1991) no. 11, pp. 1521-1527].


\bibitem{Valenzuela:2009gu} 
M.~Valenzuela,
``Higher Spin Symmetries of the Free Schrödinger Equation,''
Adv.\ Math.\ Phys.\  {\bf 2016}, 5739410 (2016)
doi:10.1155/2016/5739410
\href{http://arxiv.org/abs/arXiv:0912.0789}{arXiv:0912.0789 [hep-th]}.


\bibitem{Weinberg:1965nx} 
  S.~Weinberg,
  ``Infrared photons and gravitons,''
  Phys.\ Rev.\  {\bf 140}, B516 (1965).
  doi:10.1103/PhysRev.140.B516

 

\end{thebibliography}
\end{document}